# A Numerical Model-derived Boundary Layer Turbulence Product

Kenneth L. Pryor
Center for Satellite Applications and Research (NOAA/NESDIS)
Camp Springs, MD

**Abstract**

A suite of numerical model-derived turbulence assessment products has recently been developed by NESDIS/STAR and implemented on the World Wide Web. The suite of products is designed to indicate the potential for clear-air turbulence (CAT) resulting from Kelvin-Helmholtz instability. The existing product suite is intended to provide turbulence information to aircraft flying in the vicinity of the jet stream. Turbulence is also a frequently occurring phenomenon in the boundary layer and may pose a threat to low-flying aircraft and aircraft during the take-off and landing phases of flight. Therefore, a numerical model-derived boundary layer turbulence product is under development. The boundary layer turbulence index (TIBL) product is designed to assess the potential for turbulence in the lower troposphere, generated using National Center for Environmental Prediction (NCEP) Rapid Update Cycle (RUC)-2 model data. The index algorithm approximates boundary layer turbulent kinetic energy (TKE) by parameterizing vertical wind shear, responsible for mechanical production, and kinematic heat flux, responsible for buoyant production. The TIBL product is expected to be most effective during the afternoon hours in a regime characterized by a quasi-steady state convective mixed layer. Preliminary case studies and validation have revealed a strong correlation between the presence of TKE maxima and peak wind gust magnitude observed at the surface in evolving morning or mature afternoon mixed layers. This paper will discuss validation efforts and present case studies that demonstrate potential operational utility. Particular emphasis will be given to the coordinated use of the TIBL product with other aviation applications, such as the GOES microburst products.

## 1. Introduction

A suite of numerical model-derived turbulence assessment products has recently been developed by NESDIS/STAR and implemented on the World Wide Web. The suite of products is designed to indicate the potential for clear-air turbulence (CAT) resulting from Kelvin-Helmholtz instability (Ellrod and Knapp 1992). The prototype product, known as the deformation-vertical shear index (DVSI) or turbulence index (TI) product is currently generated utilizing operational numerical weather prediction model data. The TI algorithm consists of two terms: horizontal deformation (DEF) at a standard pressure level and total vector vertical wind shear (VWS) between two pressure levels. A recent enhancement to the TI has been the inclusion of a "divergence trend" (DVT) term to account for the time variation of divergence present in unbalanced anticyclonic jet streams (Knox et al 2006), resulting in the implementation of the D-DVSI product based on the following algorithm:

$$\text{D-DVSI} \equiv \text{DEF} \times \text{VWS} + \text{DVT} \qquad (1)$$

The existing product suite is thus intended to provide turbulence information to aircraft flying in the vicinity of the jet stream.

Turbulence is also a frequently occurring phenomenon in the boundary layer and may pose a threat to low-flying aircraft and aircraft during the take-off and landing phases of flight. Therefore, a numerical model-derived boundary layer turbulence product is under development.

The boundary layer turbulence index (TIBL) product is designed to assess the potential for turbulence in the lower troposphere, generated using National Center for Environmental Prediction (NCEP) Rapid Update Cycle (RUC)-2 model data.  Derivation of the index algorithm is based on the following assumptions and approximations as inferred from Stull (1988) and Garratt (1992):
1. Horizontal homogeneity
2. Subsidence is negligible
3. The coordinate system is aligned with the mean wind U
4. The boundary layer is sufficiently dry such that $T_v \approx T$ and $\theta_v \approx \theta$
5. $\partial\theta/\partial z \sim \partial T/\partial z$
6. $w'\theta' \sim \partial\theta/\partial z$, $w'T' \sim \partial T/\partial z$, $w'u' \sim \partial U/\partial z$, based on local closure

where $\theta$ is potential temperature, $\theta_v$ is virtual potential temperature and $T_v$ is virtual temperature. The index algorithm approximates boundary layer turbulent kinetic energy (TKE) by parameterizing the two most important production terms of the TKE budget equation as outlined in Stull (1988) and Garratt (1992):  vertical wind shear, responsible for mechanical production, and kinematic heat flux, responsible for buoyant production. Equation 2 describes the relationship between shear, buoyancy, and turbulence in boundary layer turbulence generation:

$$TKE \sim VWS + H \qquad (2)$$

where TKE represents turbulent kinetic energy, VWS represents total vector vertical wind shear between the 700 and 850-mb levels, and the quantity H represents vertical heat flux. The TKE index computation does not result in an absolute TKE value, but rather an index value that represents the relative contribution of shear and buoyancy in the generation of turbulence.  In the boundary layer, turbulence is visualized as the superimposition of many different size eddies, resulting from solar heating of the surface and the subsequent development of thermally induced eddy circulation (Sorbjan 2003).  Stull (1988) notes that the largest boundary layer eddies scale to the depth of the boundary layer and are typically the most intense, produced by solar heating, vertical wind shear, or a combination thereof.  The RUC TIBL algorithm is intended to be a mixed layer index, assessing the potential for turbulence in the boundary layer above the surface layer. Accordingly, the TIBL product parameterizes convective sources of turbulence as well as wind shear in the lower boundary layer and near the top of the boundary layer that may also contribute to turbulence generation.

Sorbjan (1989) implies that turbulent heat flux (w'T') is related to the mean vertical temperature gradient $\partial T/\partial z$. Similarly, vertical wind shear (VWS) may be represented by total vector shear $\partial V/\partial Z$ (Ellrod and Knapp 1992). In addition, the relationship between vertical wind shear and vertical momentum transfer dictates that surface wind gust magnitude will be proportional to vertical shear, and thus would be an effective measurement to quantify boundary layer turbulence.  Therefore, assuming horizontal homogeneity and substituting absolute temperature T for potential temperature $\theta$, boundary layer TKE can be approximated by equation 3, based on simplification of the TKE budget equations as given in Stull (1988) and Sorbjan (1989):

$$\partial e/\partial t \sim \partial V/\partial z + \partial T/\partial z \qquad (3)$$

The terms on the r.h.s. of equation 3 are considered to be the most important contributors to boundary layer turbulence. The second term on the r.h.s. represents the temperature lapse rate between the 700 and 850-mb levels.  Pryor (2006) identified a typical warm season mixed layer depth of 670 mb based on inspection of twenty GOES proximity soundings from the summer of 2005 over Oklahoma. It is assumed that buoyant heat flux is associated with a steep temperature lapse rate in the mixed layer that is nearly dry adiabatic.  Thus, boundary layer turbulence potential, quantified by eq. 2, may be contributed equally by significant values of VWS and H.  The TIBL product is expected to be most effective during the afternoon hours in a regime characterized by a quasi-steady state convective mixed layer. Preliminary validation has revealed a strong correlation between the presence of TKE maxima and peak wind gust magnitude observed at the surface in evolving morning or mature afternoon mixed layers.  This paper will discuss validation efforts and present case studies that demonstrate potential operational utility.  Particular emphasis will be given to the coordinated use of the TIBL product with other aviation applications, such as the GOES microburst products.

## 2. Methodology and Validation

TIBL product imagery was collected for significant wind events that occurred over the Oklahoma Panhandle region between the months of October 2006 and February 2007.  TIBL product images were generated by Man computer Interactive Data Access System (McIDAS) and then archived on an FTP server (ftp://ftp.orbit.nesdis.noaa.gov/pub/smcd/opdb/tke/).  Surface wind gusts, as observed by Oklahoma Mesonet stations in the panhandle region were compared to corresponding TKE index values, nearest in time and space.  Since $u = U + u'$ (Sorbjan 1989), variables easily measured by mesonet observing stations, comparing observations of wind speed (u) to TKE index values was the most effective means of assessing the ability of the TIBL product to represent mixed layer turbulence.

The panhandle region was selected as an area of study due to the wealth of surface observation data provided by the Oklahoma Mesonet (Brock et al. 1995), a thermodynamic environment typical of the High Plains region, and relatively homogeneous topography. The High Plains region encompasses the Oklahoma Panhandle that extends from 100° to 103° West (W) longitude. The ground elevation on the panhandle increases from near 2000 feet at 100°W longitude to nearly 5000 feet at 103°W longitude (Oklahoma Climatological Survey 1997) and the terrain is dominated by short grass prairie.  The relatively flat, treeless terrain of the Oklahoma Panhandle allows for the assumption of horizontal homogeneity. Surface wind observations, as recorded by Oklahoma Mesonet stations, were measured at a height of 10 meters (33 feet) above ground level. In addition, proximity RUC sounding profiles were collected for each significant wind event. Also, in order to assess the predictive value of the TIBL product for downburst events, data used in validation was obtained for product valid times one to three hours prior to the observed peak surface wind gusts, assuming that no change in environmental static stability and air mass characteristics between valid time and time of observed wind gusts had occurred.

For significant wind gust observations associated with convective downbursts, Next Generation Radar (NEXRAD) base reflectivity imagery (level II/III) from National Climatic Data Center (NCDC) was utilized to verify that observed wind gusts were associated with convective storm activity. NEXRAD images were generated by the NCDC Java NEXRAD Viewer (Available online at http://www.ncdc.noaa.gov/oa/radar/jnx/index.html). Another application of the NEXRAD imagery was to infer microscale physical properties of downburst-producing convective storms. Particular radar reflectivity signatures, such as the bow echo and rear-inflow notch (RIN)(Przybylinski 1995), were effective indicators of the occurrence of downbursts.

## 3. Case Studies

**14 November 2006 Downbursts**

During the afternoon of 14 November 2006, strong downbursts occurred over the Oklahoma Panhandle as observed by several Oklahoma Mesonet stations. Mid-afternoon (2100 UTC) Geostationary Operational Environmental Satellite (GOES) Hybrid Microburst Index (HMI) (Pryor 2006) product imagery indicated a favorable boundary layer thermodynamic structure and displayed high risk values (greater than or equal to 24) over the central and western panhandle. In addition, corresponding Rapid Update Cycle (RUC) boundary layer turbulence product imagery indicated elevated risk values over the same region. The combination of large GOES HMI and RUC Turbulent Kinetic Energy (TKE) index values signified the existence of a deep, dry convective mixed layer and the resultant favorability for downburst development due to sub-cloud evaporation of precipitation associated with a low-reflectivity, weakly convective rainband. Although HMI values were large, very low CAPE, as indicated in proximity sounding profiles, was more typical for warm-season dry microburst environments over the High Plains (Caracena et al 2006; Wakimoto 1985). This study will focus on downbursts observed over Cimarron County between 2200 and 2300 UTC.

Geostationary Operational Environmental Satellite (GOES) Hybrid Microburst Index (HMI) imagery and a Turbulent Kinetic Energy (TKE) Index product image based on a Rapid Update Cycle (RUC) model analysis at 2100 UTC 14 November 2006 are shown in Figure 1. The HMI product image displays high values over the central and western Oklahoma Panhandle. The corresponding TKE index product image displays maxima or "ridges" (yellow scalloped lines) over the central and western panhandle in close proximity to locations of observed downburst winds (knots, plotted in yellow over respective mesonet stations). The combination of high HMI and TKE index values indicated the presence of a deep, dry convective mixed layer with strong vertical wind shear prior to the onset of convective storm activity during the following one to three hour period. In addition to buoyancy effects, downward momentum transport by convective downdrafts was also likely a factor in the magnitude of wind gusts observed by the mesonet stations. Correlation statistics and observation data for this event are displayed in Table 1.

The first downburst-producing rainband tracked through Cimarron County between 2200 and 2300 UTC. Although radar reflectivity associated with the rainband was weak (below 15 dBZ), embedded small-scale bow echoes with associated RINs were apparent in the downburst-producing segments over Kenton and Boise City as apparent in Amarillo, Texas (KAMA) NEXRAD base reflectivity imagery in Figure 2. The strongest downburst wind gust of the event, 63 knots, displayed in Figure 3, was recorded by the Boise City mesonet station at 2225 UTC. Corresponding to the downburst at Boise City were the largest HMI value in the region of 29 and an elevated TKE index value of 18, indicated in the HMI and TKE product images at 2100 UTC, over an hour prior to downburst occurrence. The 2100 UTC HMI image displayed a rainband over southeastern Colorado that would eventually track southeastward over the Oklahoma Panhandle between 2200 UTC and 0000 UTC 15 November 2006. The location of the observed downburst winds was in close proximity to a TKE ridge, suggesting that intense turbulent mixing due to a combination of solar heating of the surface and vertical wind shear primed the boundary layer for strong convective downdraft development. A proximity RUC sounding profile at 2100 UTC in Figure 4 as well as a meteogram from Kenton in Figure 3 most effectively portrayed the favorable boundary layer thermodynamic structure that evolved during the afternoon over the western panhandle.

In addition to the "inverted-v" profile apparent in the sounding, the sounding over Kenton revealed a wind speed of 48 knots at approximately 3000 feet above ground level (AGL). The peak downburst wind gust of 48 knots recorded at 2205 UTC highlighted the role of downward horizontal momentum transfer by convective downdrafts from the mixed layer to the surface. By 2300 UTC, the favorable environment for downbursts had shifted eastward to Texas County as indicated in HMI and TKE imagery at 2300 UTC (not shown). The 2300 UTC TKE product image displayed a northeast to southwest oriented ridge over northern Texas County. Between 2300 and 0000 UTC, a series of rainbands moved eastward through Texas County in the central panhandle, producing severe downbursts at Goodwell and Hooker. Associated with the rainbands were higher radar reflectivities up to 35 dBZ (not shown). Again, in a similar manner to the downbursts that occurred earlier, bow echoes were embedded in the rainbands at the time of downburst observation.

Corfidi et al. (2004) have noted that downburst-producing mesoscale convective systems (MCSs) can occur in environments of limited surface moisture (surface dewpoints at or below 50F). In these cases, strong convective downdraft development is driven by sub-cloud evaporational cooling of precipitation resulting from a combination of steep low to mid level lapse rates and large dewpoint depressions. Since the HMI incorporates the low to mid level temperature lapse rate and dewpoint depression difference (Pryor 2006), it is expected that downburst activity will be maximized in regions of large HMI values. It was also noted that moderate to strong vertical wind shear is instrumental in strengthening and deepening low-level system relative inflow, and hence contributing to MCS sustenance. Thus, in this case, the development of cold convective downdrafts in the presence of vertical wind shear was an important factor in downburst generation and system maintenance, demonstrating the value of the GOES HMI and RUC TIBL products in the short-term forecasting of downburst potential.

**Non-convective High Winds 24 February 2007**

During the morning of 24 February 2007, non-convective high winds were observed over the western Oklahoma Panhandle behind a strong cold front. The late morning (1700 UTC) Rapid Update Cycle (RUC) boundary layer turbulence product (Turbulent Kinetic Energy, TKE) image indicated a maximum in index values over southeastern Colorado and northeastern New Mexico with a large gradient in TKE index values over Cimarron County, Oklahoma. A peak wind gust of 59 knots was observed by Kenton and Boise City, Oklahoma mesonet stations between 1700 and 1800 UTC. During this high wind event, elevated RUC TKE index values signified the existence of a mixed layer with strong vertical wind shear, and the resultant favorability for high surface wind gusts due to downward horizontal momentum transport. A corresponding RUC model-derived sounding reflected favorable conditions for strong and gusty surface winds displaying a shallow "inverted-v" profile with significant increase in wind speed with height in the mixed layer.

Figure 5 displays a Turbulent Kinetic Energy (TKE) Index product image based on a Rapid Update Cycle (RUC) model analysis at 1700 UTC 24 February 2007. Surface observation analysis indicated that a strong cold front tracked eastward through the Oklahoma Panhandle between 1100 and 1700 UTC. After 1200 UTC, as displayed in the Kenton meteogram in Figure 6, the surface wind speed increased sharply with several gusts in excess of 50 knots at Kenton between 1200 and 1400 UTC. Winds diminished slightly between 1400 and 1600 UTC. However, by 1600 UTC (not shown), TKE index product imagery displayed a maximum over southeastern Colorado and northeastern New Mexico with a large gradient in index values over Cimarron County, Oklahoma. Between 1600 and 1700 UTC, TKE index values increased from 22 to 26 at Kenton. This increase in index values most likely was a consequence of the morning evolution of the mixed layer and resultant increase in boundary layer depth and eddy circulation as well as an increase in vertical wind shear. Figure 7, the 1700 UTC RUC sounding at Kenton, reflected favorable conditions for strong and gusty surface winds by displaying a shallow "inverted-v" profile with a significant increase in wind speed with height in the mixed layer. In addition, the sounding indicated a wind speed near the top of the mixed layer near 50 knots. At 1710 UTC, a wind gust of 59 knots was recorded at the Kenton mesonet station followed by a gust of 59 knots at Boise City at 1755 UTC. Downward momentum transport by intense eddy circulations was likely a factor in the magnitude of observed surface wind gusts in a shear-driven boundary layer.

**4. Summary and Conclusions**

A boundary layer turbulence index (TIBL) product has been developed to assess the potential for turbulence in the lower troposphere, generated using RUC-2 numerical model data. The index algorithm approximates boundary layer turbulent kinetic energy by parameterizing vertical wind shear, responsible for mechanical production of TKE, and kinematic heat flux, parameterized by the vertical temperature lapse rate $\partial T/\partial z$ and responsible for buoyant production of TKE. Validation for the TIBL product has been conducted for selected convective and non-convective wind events. This paper presented studies of two significant wind events during November 2006 and February 2007 over the Oklahoma Panhandle region. It was found that for the 14 November downburst event, the observation of strong surface wind gusts occurred in close proximity to local maxima or "ridges" of TKE. This case also demonstrated a close correspondence between TKE index values and GOES sounder-derived microburst risk values for downburst occurrence. In contrast, the 24 February high wind event demonstrated the effectiveness of the TIBL product in quantifying the evolution of turbulence in an evolving morning mixed layer. Apparent in this study was a strong correlation between a TKE maximum and the magnitude of peak surface wind gusts (U + u') observed at Kenton and Boise City. In a similar manner to the convective wind case, peak surface wind gusts were observed in close proximity to a TKE ridge. Based on the favorable results highlighted in the case studies, the RUC TIBL product should have operational utility in assessing hazards to low-flying aircraft. Further validation during the warm season will be necessary to optimize the capability of the TIBL product in the evaluation of turbulence risk to aviation.

**Acknowledgements**

The author thanks Mr. Derek Arndt (Oklahoma Climatological Survey) and the Oklahoma Mesonet for the surface weather observation data used in this research effort. RUC model sounding profiles were provided by NOAA website **rucsoundings.noaa.gov**.

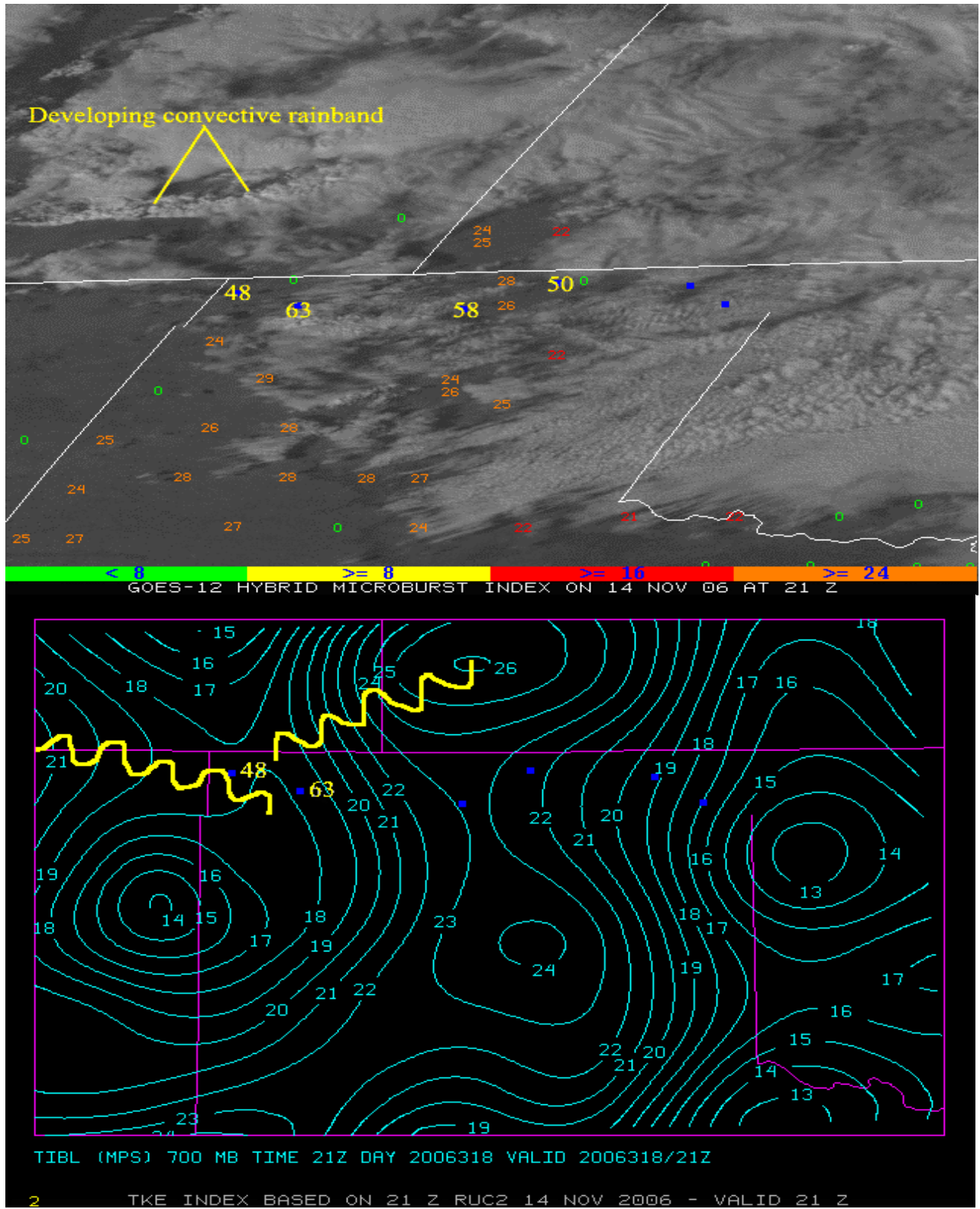

Figure 1. GOES HMI (top) and RUC TIBL (bottom) images at 2100 UTC 14 November 2006.

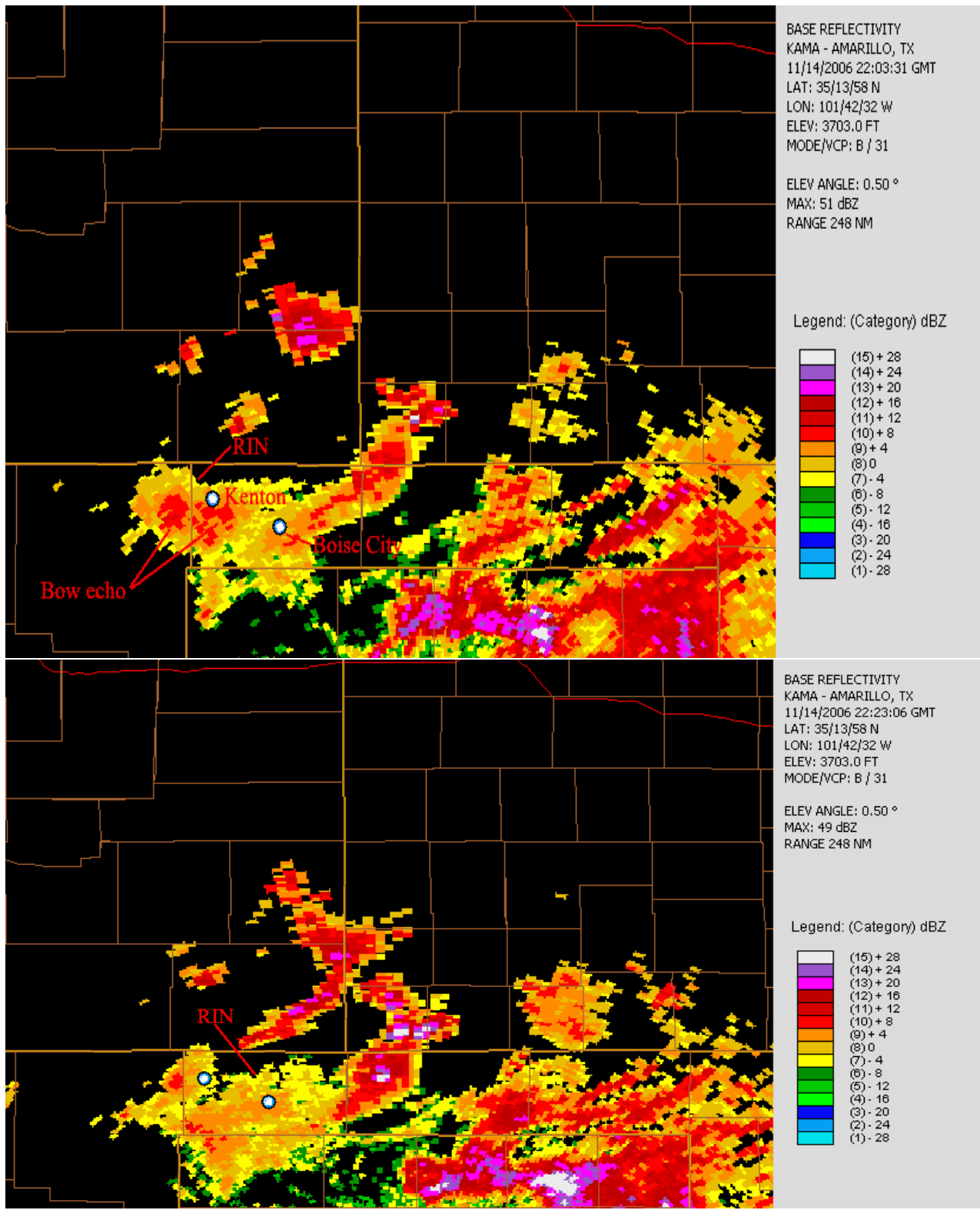

Figure 2.  NEXRAD reflectivity at 2203 UTC (top) and 2223 UTC (bottom) 14 November 2006.

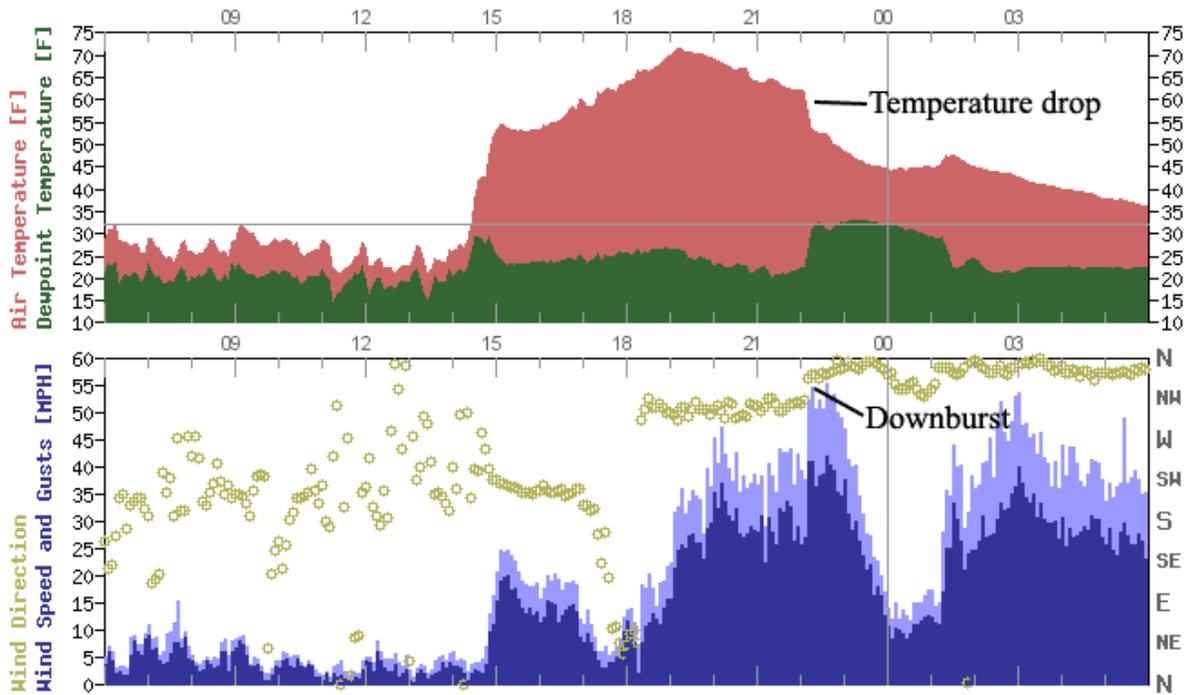
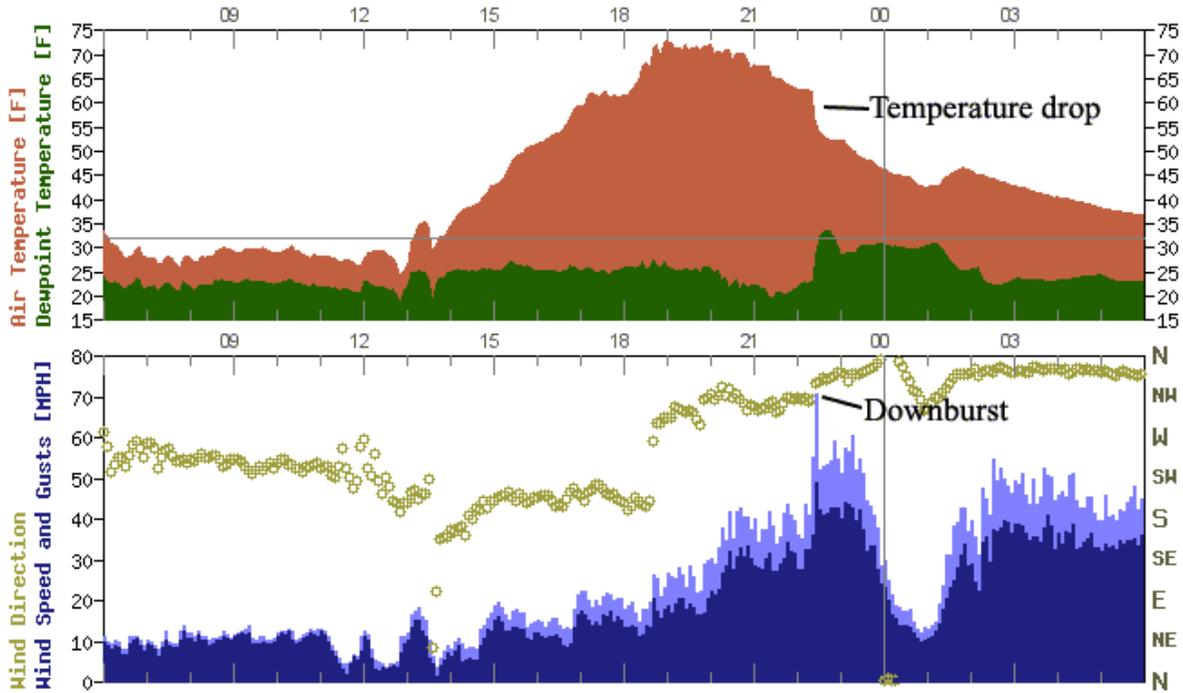

Figure 3.  Oklahoma Mesonet meteograms at Kenton (top) and Boise City (bottom).

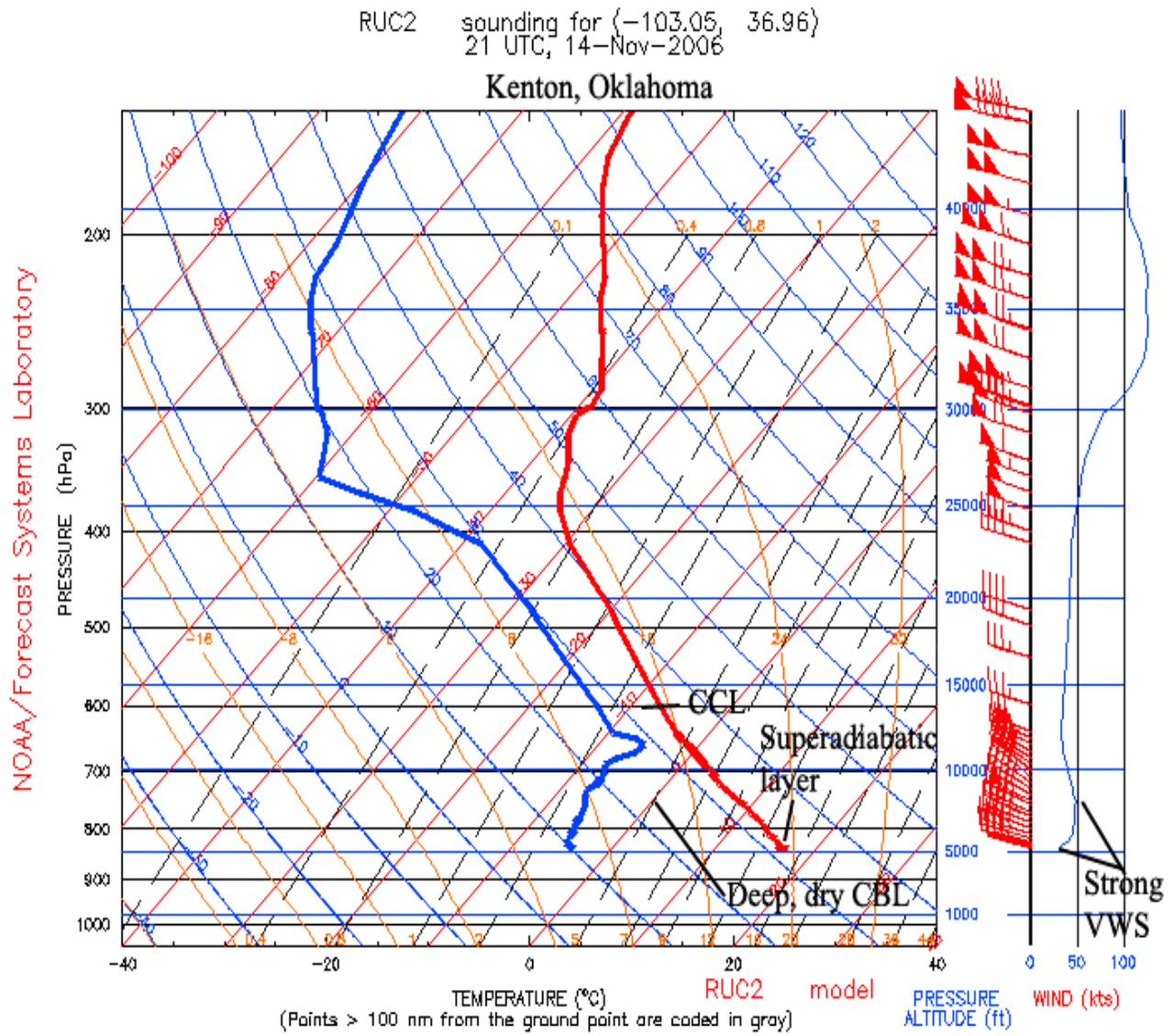

Figure 4. RUC-2 sounding at Kenton at 2100 UTC 14 November 2006.

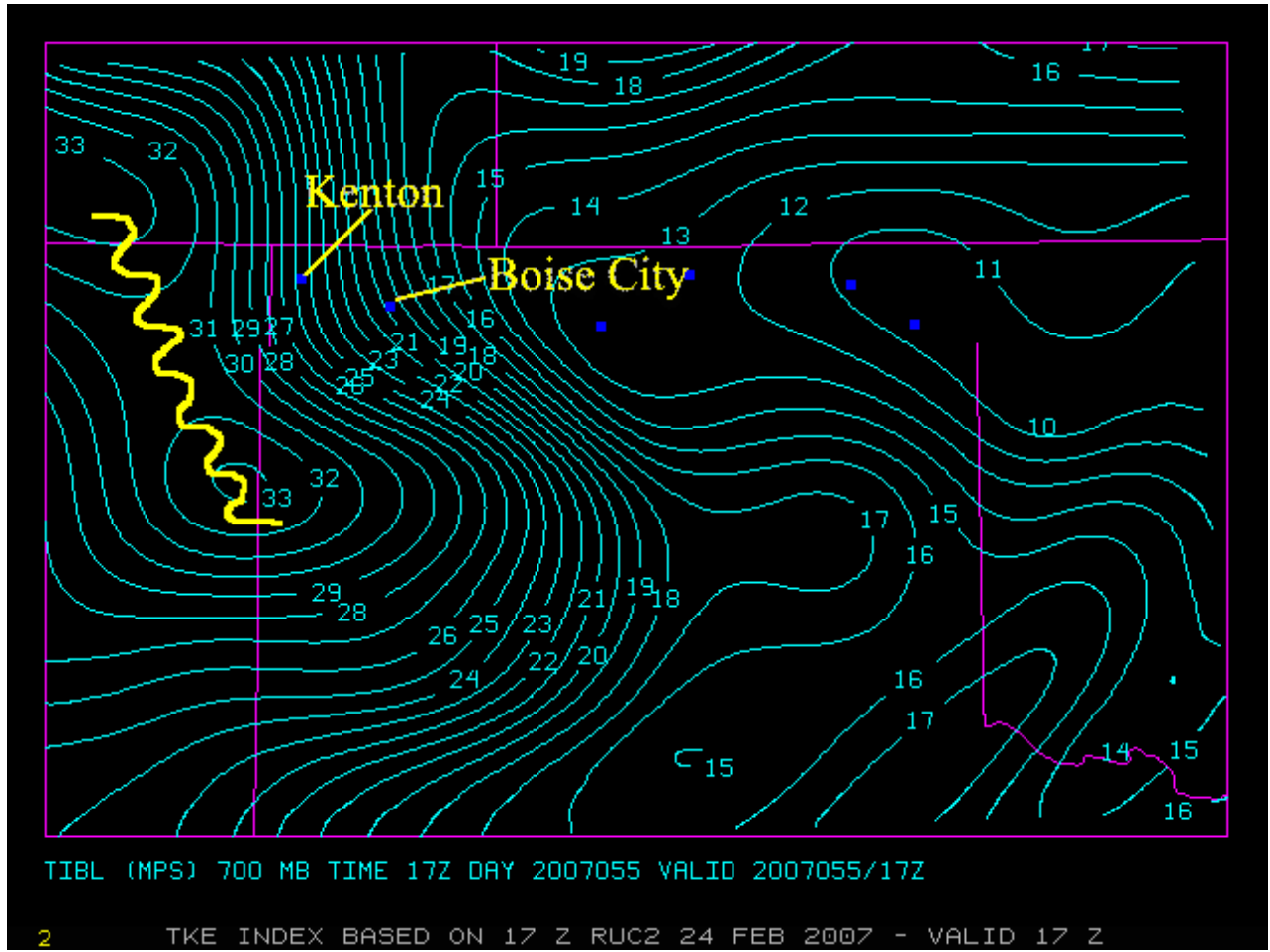

Figure 5. RUC TIBL image at 1700 UTC 24 February 2007.

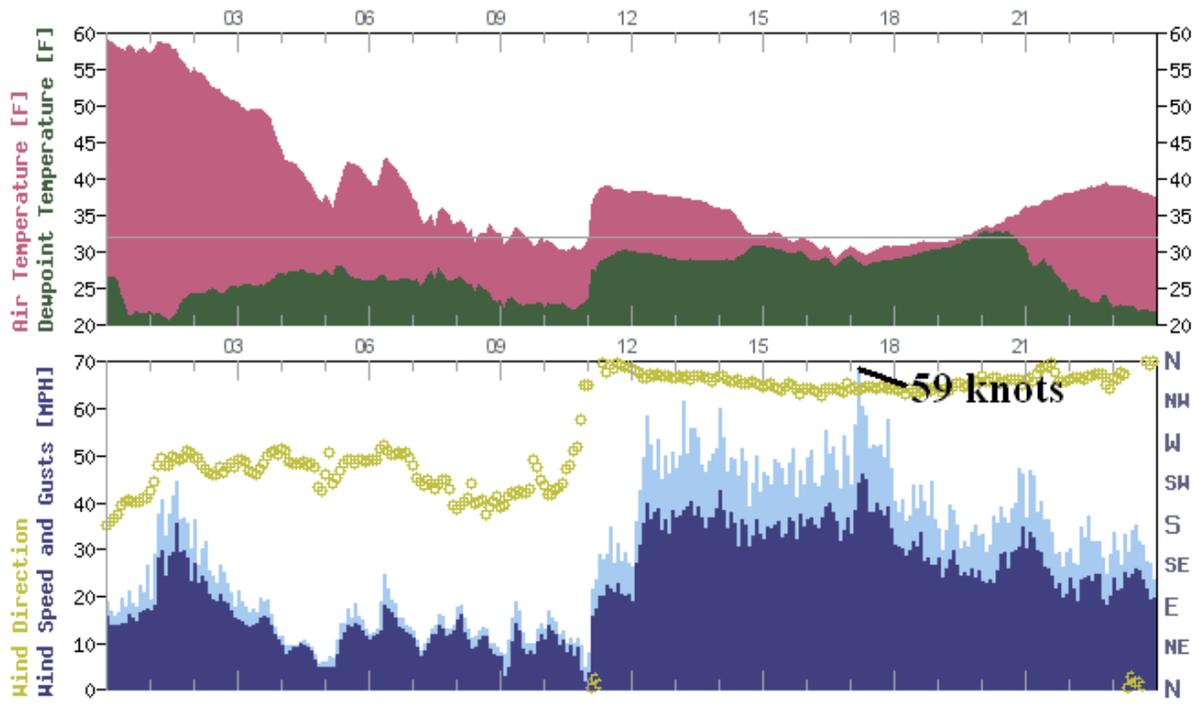

Figure 6. 24 February 2007 Oklahoma Mesonet meteogram at Kenton.

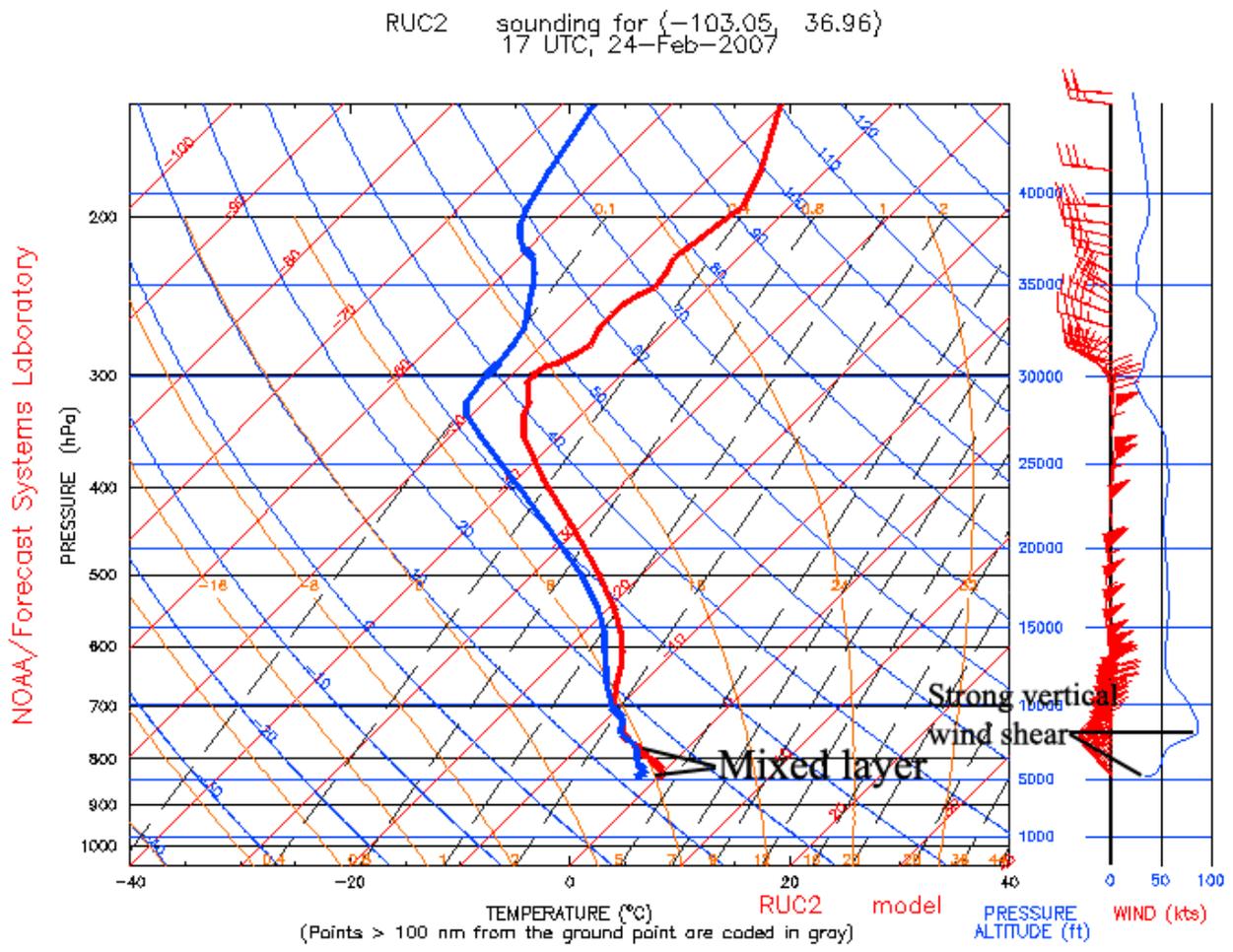

Figure 7. RUC-2 sounding at Kenton at 1700 UTC 24 February 2007.

**14 Nov 06**

| | | | | | |
|---|---|---|---|---|---|
| WMSI to measured wind: | 0.64 | WMSI to HMI: | 0.46 | Mean HMI: | 25.75 |
| HMI to measured wind: | 0.96 | | | Mean WMSI: | -1 |
| TKE to measured wind: | 0.48 | No. of events: | 4 | Mean Wind Speed: | 55 |
| | | | | Mean TKE | 17 |

| Time | Measured Wind Speed kt | Location | GOES-12 HMI | RUC TKE | GOES-12 WMSI | DD(F) |
|---|---|---|---|---|---|---|
| 22:10 | 48 | Kenton | 24 | 18 | -4 | 40 |
| 22:25 | 63 | Boise City | 29 | 18 | 0 | 43 |
| 23:45 | 58 | Goodwell | 26 | 19 | 0 | 35 |
| 0:00 | 50 | Hooker | 24 | 14 | 0 | 32 |

Table1. Correlation statistics and observation data for 14 November 2006 downburst event.